\documentstyle[12pt]{article}
\makeatletter
\newdimen\normalarrayskip              
\newdimen\minarrayskip                 
\normalarrayskip\baselineskip
\minarrayskip\jot
\newif\ifold             \oldfalse
\newif\ifdisplayarray    \displayarraytrue
\newif\ifbigarray        \bigarraytrue
\def\arraymode{\ifold\relax\else\ifdisplayarray\displaystyle\else\relax\fi\fi}
\def\eqnumphantom{\phantom{(\theequation)}}     
\def\@arrayskip{\ifold\baselineskip\z@\lineskip\z@\else\ifbigarray
     \baselineskip\normalarrayskip\lineskip\minarrayskip
     \else
     \baselineskip\z@\lineskip\z@\fi\fi}
\def\@arrayclassz{\ifcase \@lastchclass \@acolampacol \or
\@ampacol \or \or \or \@addamp \or
   \@acolampacol \or \@firstampfalse \@acol \fi
\edef\@preamble{\@preamble
  \ifcase \@chnum
     \hfil$\relax\arraymode\@sharp$\hfil
     \or $\relax\arraymode\@sharp$\hfil
     \or \hfil$\relax\arraymode\@sharp$\fi}}
\def\@array[#1]#2{\setbox\@arstrutbox=\hbox{\vrule
     height\arraystretch \ht\strutbox
     depth\arraystretch \dp\strutbox
     width\z@}\@mkpream{#2}\edef\@preamble{\halign \noexpand\@halignto
\bgroup \tabskip\z@ \@arstrut \@preamble \tabskip\z@ \cr}%
\let\@startpbox\@@startpbox \let\@endpbox\@@endpbox
  \if #1t\vtop \else \if#1b\vbox \else \vcenter \fi\fi
  \bgroup \let\par\relax
  \let\@sharp##\let\protect\relax
  \@arrayskip\@preamble}
\def\eqnarray{\stepcounter{equation}%
              \let\@currentlabel=\theequation
              \global\@eqnswtrue
              \global\@eqcnt\z@
              \tabskip\@centering
              \let\\=\@eqncr
              $$%
 \halign to \displaywidth\bgroup
    \eqnumphantom\@eqnsel\hskip\@centering
    $\displaystyle \tabskip\z@ {##}$%
    &\global\@eqcnt\@ne \hskip 2\arraycolsep
         \hfil$\arraymode{##}$\hfil
    &\global\@eqcnt\tw@ \hskip 2\arraycolsep
         $\displaystyle\tabskip\z@{##}$\hfil
         \tabskip\@centering
    &{##}\tabskip\z@\cr}
\newenvironment{marray}{\begin{equation}\begin{array}}%
{\end{array}\end{equation}}
\newenvironment{carray}{\begin{equation}\begin{array}{rcl}}%
{\end{array}\end{equation}}
\def\be{\@ifnextchar[{\def\ee{\end{equation}}\begin{equation}\l@b}%
{\def\ee{$$}$$}}
\def\l@b[#1]{\label{#1}}
\def\ba{\@ifnextchar[{\def\ee{\end{carray}}\begin{carray}\l@b}%
{\def\ee{\end{array}$$}$$\begin{array}{rcl}}}
\def\barray#1{\@ifnextchar[{\def\ee{\end{marray}}\begin{marray}{#1}\l@b}%
{\def\ee{\end{array}$$}$$\begin{array}{#1}}}
\def\herring{\@ifnextchar[{\@herring}{\@herring[\vcenter]}}
\def\@herring[#1]#2{\begingroup
\def\*{\\ \>}
\topsep0pt
\partopsep0pt
\def\tabbing{\lineskip\jot \lineskiplimit\jot
     \let\>\@rtab\let\<\@ltab\let\=\@settab
     \let\+\@tabplus\let\-\@tabminus\let\`\@tabrj\let\'\@tablab
     \let\\=\@tabcr
     \global\@hightab\@firsttab
     \global\@nxttabmar\@firsttab
     \dimen\@firsttab\@totalleftmargin
     \global\@tabpush0 \global\@rjfieldfalse
     \trivlist \item[]\if@minipage\else\vskip\parskip\fi
     \setbox\@tabfbox\hbox{\rlap{\indent\hskip\@totalleftmargin
       \the\everypar}}\def\@itemfudge{\box\@tabfbox}\@startline\ignorespaces}
\def\@startfield{\global\setbox\@curfield\hbox
                    \bgroup$\displaystyle}%
\def\@stopfield{$\egroup}%
#1{\begin{tabbing}#2\end{tabbing}}\endgroup}
\def\eq#1{(\ref{#1})}
\def\theequation{\thesection.\arabic{equation}}
\@addtoreset{equation}{section}%
\def\@cite#1#2{\hbox{ [#1\if@tempswa ,#2\fi]}}
\def\@citex[#1]#2{\if@filesw\immediate\write\@auxout{\string\citation{#2}}\fi
  \def\@citea{}\@cite{\@for\@citeb:=#2\do
    {\@citea\def\@citea{,\penalty\@m}\@ifundefined  
       {b@\@citeb}{{\bf ?}\@warning
       {Citation `\@citeb' on page \thepage \space undefined}}%
\hbox{\csname b@\@citeb\endcsname}}}{#1}}
\def\@sect#1#2#3#4#5#6[#7]#8{\ifnum #2>\c@secnumdepth
     \def\@svsec{}\else
     \refstepcounter{#1}\edef\@svsec{\csname the#1\endcsname.%
     \hskip 0.8em }\fi
     \@tempskipa #5\relax
      \ifdim \@tempskipa>\z@
        \begingroup #6\relax
          \@hangfrom{\hskip #3\relax\@svsec}{\interlinepenalty \@M #8\par}%
        \endgroup
       \csname #1mark\endcsname{#7}\addcontentsline
         {toc}{#1}{\ifnum #2>\c@secnumdepth \else
                      \protect\numberline{\csname the#1\endcsname}\fi
                    #7}\else
        \def\@svsechd{#6\hskip #3\@svsec #8\csname #1mark\endcsname
                      {#7}\addcontentsline
                           {toc}{#1}{\ifnum #2>\c@secnumdepth \else
                             \protect\numberline{\csname the#1\endcsname}\fi
                       #7}}\fi
     \@xsect{#5}}

\makeatother
\begin{document}
\newcommand {\ignore}[1]{}
\newcommand{\nota}[1]{\makebox[0pt]{\,\,\,\,\,/}#1}
\newcommand{\notp}[1]{\makebox[0pt]{\,\,\,\,/}#1}
\newcommand{\braket}[1]{\mbox{$<$}#1\mbox{$>$}}
\newcommand{\Frac}[2]{\frac{\displaystyle #1}{\displaystyle #2}}
\renewcommand{\arraystretch}{1.5}
\newcommand{\noi}{\noindent}
\newcommand{\bc}{\begin{center}}
\newcommand{\ec}{\end{center}}
\newcommand{\epm}{e^+e^-}
\def\ifmath#1{\relax\ifmmode #1\else $#1$\fi}
%
\def\half{\ifmath{{\textstyle{1 \over 2}}}}
\def\quarter{\ifmath{{\textstyle{1 \over 4}}}}
\def\3quarter{{\textstyle{3 \over 4}}}
\def\third{\ifmath{{\textstyle{1 \over 3}}}}
\def\twothirds{{\textstyle{2 \over 3}}}
\def\fourth{\ifmath{{\textstyle{1\over 4}}}}
\def\sqrthalf{\ifmath{{\textstyle{1\over\sqrt2}}}}
\def\halfsqrthalf{\ifmath{{\textstyle{1\over2\sqrt2}}}}
\def\nl{\nextline}
\def\cl{\centerline}
\def\vs{\vskip}
\def\hs{\hskip}
\def\ss{\smallskip}
\def\ms{\medskip}
\def\bs{\bigskip}
\def\br{\break}
\def\ra{\rightarrow}
\def\Ra{\Rightarrow}
\def\us{\undertext}
\def\HB{\hfill\break}
\overfullrule 0pt
\def\lf{\leaders\hbox to 1em{\hss.\hss}\hfill}
\def\ZP{$Z^\prime$ }
\def\21{$SU(2) \ot U(1)$}
\def\321{$SU(3) \ot SU(2) \ot U(1)$}
\def\ne{\hbox{$\nu_e$ }}
\def\nm{\hbox{$\nu_\mu$ }}
\def\nt{\hbox{$\nu_\tau$ }}
\def\ns{\hbox{$\nu_{sterile}$ }}
\def\nx{\hbox{$\nu_x$ }}
\def\Nt{\hbox{$N_\tau$ }}
\def\ns{\hbox{$\nu_S$ }}
\def\nr{\hbox{$\nu_R$ }}
\def\O{\hbox{$\cal O$ }}
\def\L{\hbox{$\cal L$ }}
\def\mne{\hbox{$m_{\nu_e}$ }}
\def\mnm{\hbox{$m_{\nu_\mu}$ }}
\def\mnt{\hbox{$m_{\nu_\tau}$ }}
\def\mq{\hbox{$m_{q}$}}
\def\ml{\hbox{$m_{l}$}}
\def\mup{\hbox{$m_{u}$}}
\def\md{\hbox{$m_{d}$}}
\def\ie{\hbox{\it i.e., }}        \def\etc{\hbox{\it etc. }}
\def\eg{\hbox{\it e.g., }}        \def\cf{\hbox{\it cf.}}
\def\etal{\hbox{\it et al., }}
\def\H{\hbox{Higgs }}
\def\nhl{\hbox{neutral heavy lepton }}
\def\NHL{\hbox{Neutral Heavy Lepton }}
\def\nhls{\hbox{neutral heavy leptons }}
\def\NHLs{\hbox{Neutral Heavy Leptons }}
\def\Hp{\hbox{Higgs. }}
\def\nhlp{\hbox{neutral heavy lepton. }}
\def\NHLp{\hbox{Neutral Heavy Lepton. }}
\def\nhlsp{\hbox{neutral heavy leptons. }}
\def\NHLsp{\hbox{Neutral Heavy Leptons. }}
\def\nv{\hbox{non-vanishing }}
\def\nvp{\hbox{non-vanishing. }}
\def\nvc{\hbox{non-vanishing, }}
\def\fd{\hbox{field }}
\def\Fd{\hbox{Field }}
\def\fds{\hbox{fields }}
\def\eig{\hbox{eigenstate }}
\def\Def{\hbox{Definition }}
\def\defd{\hbox{defined }}
\def\wf{\hbox{wave-function }}
\def\wfs{\hbox{wave-functions }}
\def\wfp{\hbox{wave-function. }}
\def\wfc{\hbox{wave-function, }}
\def\meig{\hbox{mass-eigenstate }}
\def\meigs{\hbox{mass-eigenstates }}
\def\meigc{\hbox{mass-eigenstate, }}
\def\meigp{\hbox{mass-eigenstate. }}
\def\Sst{\hbox{Superstring }}
\def\sst{\hbox{superstring }}
\def\susym{\hbox{supersymmetry }}
\def\Susym{\hbox{Supersymmetry }}
\def\susymp{\hbox{supersymmetry. }}
\def\Susymp{\hbox{Supersymmetry. }}
\def\susymc{\hbox{supersymmetry, }}
\def\Susymc{\hbox{Supersymmetry, }}
\def\susy{\hbox{supersymmetric }}
\def\Susy{\hbox{Supersymmetric }}
\def\susyp{\hbox{supersymmetric. }}
\def\Susyp{\hbox{Supersymmetric. }}
\def\susyc{\hbox{supersymmetric, }}
\def\Susyc{\hbox{Supersymmetric, }}
\def\sym{\hbox{symmetry }}
\def\sym{\hbox{symmetry }}
\def\symp{\hbox{symmetry. }}
\def\symp{\hbox{symmetry. }}
\def\symc{\hbox{symmetry, }}
\def\symc{\hbox{symmetry, }}
\def\sy{\hbox{symmetric }}
\def\sy{\hbox{symmetric }}
\def\syp{\hbox{symmetric. }}
\def\syp{\hbox{symmetric. }}
\def\syc{\hbox{symmetric, }}
\def\syc{\hbox{symmetric, }}
\def\lh{\hbox{left-handed }}
\def\Lh{\hbox{Left-handed }}
\def\LH{\hbox{Left-Handed }}
\def\lhc{\hbox{left-handed, }}
\def\lhp{\hbox{left-handed. }}
\def\rh{\hbox{right-handed }}
\def\Rh{\hbox{Right-handed }}
\def\RH{\hbox{Right-Handed }}
\def\Rhc{\hbox{Right-handed, }}
\def\rhp{\hbox{right-handed. }}
\def\sol{\hbox{solution }}
\def\sols{\hbox{solutions }}
\def\solc{\hbox{solution. }}
\def\solp{\hbox{solution. }}
\def\rep{\hbox{representation }}
\def\reps{\hbox{representations }}
\def\repc{\hbox{representation, }}
\def\repp{\hbox{representation. }}
\def\ew{\hbox{electro-weak }}
\def\Ew{\hbox{Electro-weak }}
\def\Em{\hbox{Electromagnetic }}
\def\me{\hbox{matrix element }}
\def\Me{\hbox{Matrix element }}
\def\nad{\hbox{non-adiabatic }}
\def\ad{\hbox{adiabatic }}
\def\Nad{\hbox{Non-adiabatic }}
\def\Ad{\hbox{Adiabatic }}
\def\lfvg{\hbox{lepton flavour violating }}
\def\lfv{\hbox{lepton flavour violation }}
\def\LFV{\hbox{Lepton Flavour Violation }}
\def\IVB{\hbox{Intermediate Vector Boson }}
\def\IVBs{\hbox{Intermediate Vector Bosons }}
\def\ivb{\hbox{intermediate vector boson }}
\def\IGB{\hbox{Intermediate Gauge Boson }}
\def\igb{\hbox{intermediate gauge boson }}
\def\igbs{\hbox{intermediate gauge bosons }}
\def\TLNV{\hbox{Total Lepton Number Violation }}
\def\tlnv{\hbox{total lepton number violation }}
\def\tlnvg{\hbox{total lepton number violating }}
\def\TLN{\hbox{Total Lepton Number }}
\def\tln{\hbox{total lepton number }}
\def\df{\hbox{degrees of freedom }}
\def\Df{\hbox{Degrees of freedom }}
\def\DF{\hbox{Degrees of Freedom }}
\def\gau{\hbox{gauge }}
\def\he{\hbox{high energy }}
\def\HE{\hbox{High Energy }}
\def\sn{\hbox{supernova }}
\def\Sn{\hbox{Supernova }}
\def\Gau{\hbox{Gauge }}
\def\Conf{\hbox{Conference }}
\def\inv{\hbox{invariant }}
\def\stan{\hbox{standard }}
\def\ST{\hbox{Standard Theory }}
\def\st{\hbox{standard theory }}
\def\Lor{\hbox{Lorentz }}
\def\transf{\hbox{transformation }}
\def\expt{\hbox{experiment }}
\def\expts{\hbox{experiments }}
\def\lab{\hbox{laboratory }}
\def\br{\hbox{branching ratio }}
\def\BR{\hbox{Branching Ratio }}
\def\brs{\hbox{branching ratios }}
\def\BRs{\hbox{Branching Ratios }}
\def\Lag{\hbox{Lagrangian }}
\def\wi{\hbox{weak interaction }}
\def\wis{\hbox{weak interactions }}
\def\Dir{\hbox{Dirac }}
\def\Maj{\hbox{Majorana }}
\def\cosm{\hbox{cosmology }}
\def\astro{\hbox{astrophysics }}
\def\Cosm{\hbox{Cosmology }}
\def\Astro{\hbox{Astrophysics }}
\def\mos{\hbox{models }}
\def\Mos{\hbox{Models }}
\def\sm{\hbox{standard model }}
\def\SM{\hbox{Standard Model }}
\def\CC{\hbox{Charged Current }}
\def\cc{\hbox{charged current }}
\def\NC{\hbox{Neutral Current }}
\def\neu{\hbox{neutrino }}
\def\sa{\hbox{such as }}
\def\neuless{\hbox{neutrinoless }}
\def\Neuless{\hbox{Neutrinoless }}
\def\Neus{\hbox{Neutrinos }}
\def\neus{\hbox{neutrinos }}
\def\Neu{\hbox{Neutrino }}
\def\phys{\hbox{physics }}
\def\Phys{\hbox{Physics }}
\def\Neuphys{\hbox{Neutrino physics }}
\def\neuphys{\hbox{neutrino physics }}
\def\NeuPhys{\hbox{Neutrino Physics }}
\def\Neuphysp{\hbox{Neutrino physics. }}
\def\Neuphysc{\hbox{Neutrino physics, }}
\def\osc{\hbox{oscillation }}
\def\oscs{\hbox{oscillations }}
\def\Osc{\hbox{Oscillation }}
\def\Oscs{\hbox{Oscillations }}
%
%
\def\mec{\hbox{matrix element, }}
\def\lfvgc{\hbox{lepton flavour violating, }}
\def\lfvc{\hbox{lepton flavour violation, }}
\def\LFVc{\hbox{Lepton Flavour Violation, }}
\def\IVBc{\hbox{Intermediate Vector Boson, }}
\def\IVBsc{\hbox{Intermediate Vector Bosons, }}
\def\ivbc{\hbox{intermediate vector boson, }}
\def\IGBc{\hbox{Intermediate Gauge Boson, }}
\def\igbc{\hbox{intermediate gauge boson, }}
\def\igbsc{\hbox{intermediate gauge bosons, }}
\def\brc{\hbox{branching ratio, }}
\def\BRc{\hbox{Branching Ratio, }}
\def\brsc{\hbox{branching ratios, }}
\def\BRsc{\hbox{Branching Ratios, }}
\def\TLNVc{\hbox{Total Lepton Number Violation, }}
\def\tlnvc{\hbox{total lepton number violation, }}
\def\tlnvgc{\hbox{total lepton number violating, }}
\def\TLNc{\hbox{Total Lepton Number, }}
\def\tlnc{\hbox{total lepton number, }}
\def\dfc{\hbox{degrees of freedom, }}
\def\Dfc{\hbox{Degrees of freedom, }}
\def\DFc{\hbox{Degrees of Freedom, }}
\def\gauc{\hbox{gauge, }}
\def\hec{\hbox{high energy, }}
\def\HEc{\hbox{High Energy, }}
\def\snc{\hbox{supernova, }}
\def\Snc{\hbox{Supernova, }}
\def\Gauc{\hbox{Gauge, }}
\def\Confc{\hbox{Conference, }}
\def\invc{\hbox{invariant, }}
\def\STc{\hbox{Standard Theory, }}
\def\stc{\hbox{standard theory, }}
\def\stanc{\hbox{standard, }}
\def\Stanc{\hbox{Standard, }}
\def\Lorc{\hbox{Lorentz, }}
\def\transfc{\hbox{transformation, }}
\def\exptc{\hbox{experiment, }}
\def\exptsc{\hbox{experiments, }}
\def\labc{\hbox{laboratory, }}
\def\Lagc{\hbox{Lagrangian, }}
\def\wic{\hbox{weak interaction, }}
\def\wisc{\hbox{weak interactions, }}
\def\Dirc{\hbox{Dirac, }}
\def\Majc{\hbox{Majorana, }}
\def\moc{\hbox{model, }}
\def\Moc{\hbox{Model, }}
\def\cosmc{\hbox{cosmology, }}
\def\astroc{\hbox{astrophysics, }}
\def\Cosmc{\hbox{Cosmology, }}
\def\Astroc{\hbox{Astrophysics, }}
\def\mosc{\hbox{models, }}
\def\Mosc{\hbox{Models, }}
\def\smc{\hbox{standard model, }}
\def\SMc{\hbox{Standard Model, }}
\def\CCc{\hbox{Charged Current, }}
\def\NCc{\hbox{Neutral Current, }}
\def\ccc{\hbox{charged current, }}
\def\nc{\hbox{neutral current }}
\def\ncc{\hbox{neutral current, }}
\def\neuc{\hbox{neutrino, }}
\def\physc{\hbox{physics, }}
\def\Physc{\hbox{Physics, }}
\def\oscc{\hbox{oscillation, }}
\def\oscsc{\hbox{oscillations, }}
\def\Oscc{\hbox{Oscillation, }}
\def\Oscsc{\hbox{Oscillations, }}
\def\neusc{\hbox{neutrinos, }}
\def\Neuc{\hbox{Neutrino, }}
\def\Neusc{\hbox{Neutrinos, }}
%
%
\def\mep{\hbox{matrix element. }}
\def\nadp{\hbox{non-adiabatic. }}
\def\adp{\hbox{adiabatic. }}
\def\Nadp{\hbox{Non-adiabatic. }}
\def\Adp{\hbox{Adiabatic. }}
\def\brp{\hbox{branching ratio. }}
\def\BRp{\hbox{Branching Ratio. }}
\def\brsp{\hbox{branching ratios. }}
\def\BRsp{\hbox{Branching Ratios. }}
\def\lfvgp{\hbox{lepton flavour violating. }}
\def\lfvp{\hbox{lepton flavour violation. }}
\def\LFVp{\hbox{Lepton Flavour Violation. }}
\def\IVBp{\hbox{Intermediate Vector Boson. }}
\def\IVBsp{\hbox{Intermediate Vector Bosons. }}
\def\ivbp{\hbox{intermediate vector boson. }}
\def\IGBp{\hbox{Intermediate Gauge Boson. }}
\def\igbp{\hbox{intermediate gauge boson. }}
\def\igbsp{\hbox{intermediate gauge bosons. }}
\def\TLNVp{\hbox{Total Lepton Number Violation. }}
\def\tlnvp{\hbox{total lepton number violation. }}
\def\tlnvgp{\hbox{total lepton number violating. }}
\def\TLNp{\hbox{Total Lepton Number. }}
\def\tlnp{\hbox{total lepton number. }}
\def\dfp{\hbox{degrees of freedom. }}
\def\Dfp{\hbox{Degrees of freedom. }}
\def\DFp{\hbox{Degrees of Freedom. }}
\def\gaup{\hbox{gauge. }}
\def\hep{\hbox{high energy. }}
\def\HEp{\hbox{High Energy. }}
\def\snp{\hbox{supernova. }}
\def\Snp{\hbox{Supernova. }}
\def\Gaup{\hbox{Gauge. }}
\def\Confp{\hbox{Conference. }}
\def\invp{\hbox{invariant. }}
\def\stanp{\hbox{standard. }}
\def\Stanp{\hbox{Standard. }}
\def\Lorp{\hbox{Lorentz. }}
\def\transfp{\hbox{transformation. }}
\def\exptp{\hbox{experiment. }}
\def\exptsp{\hbox{experiments. }}
\def\labp{\hbox{laboratory. }}
\def\Lagp{\hbox{Lagrangian. }}
\def\wip{\hbox{weak interaction. }}
\def\wisp{\hbox{weak interactions. }}
\def\Dirp{\hbox{Dirac. }}
\def\Majp{\hbox{Majorana. }}
\def\mop{\hbox{model. }}
\def\Mop{\hbox{Model. }}
\def\cosmp{\hbox{cosmology. }}
\def\astrop{\hbox{astrophysics. }}
\def\Cosmp{\hbox{Cosmology. }}
\def\Astrop{\hbox{Astrophysics. }}
\def\mosp{\hbox{models. }}
\def\Mosp{\hbox{Models. }}
\def\smp{\hbox{standard model. }}
\def\SMp{\hbox{Standard Model. }}
\def\CCp{\hbox{Charged Current. }}
\def\NCp{\hbox{Neutral Current. }}
\def\NCsp{\hbox{Neutral Currents. }}
\def\ncsp{\hbox{neutral currents. }}
\def\ncs{\hbox{neutral currents }}
\def\neup{\hbox{neutrino. }}
\def\physp{\hbox{physics. }}
\def\Physp{\hbox{Physics. }}
\def\oscp{\hbox{oscillation. }}
\def\oscsp{\hbox{oscillations. }}
\def\Oscp{\hbox{Oscillation. }}
\def\Oscsp{\hbox{Oscillations. }}
\def\neusp{\hbox{neutrinos. }}
\def\Neup{\hbox{Neutrino. }}
\def\Neusp{\hbox{Neutrinos. }}
\def\CPv{\hbox{CP violation }}
\def\CPvp{\hbox{CP violation. }}
\def\CPvc{\hbox{CP violation, }}
\def\dash{\hbox{---}}
\def\c{\mathop{\cos \theta }}
\def\s{\mathop{\sin \theta }}
\def\cok{\mathop{\rm cok}}
\def\tr{\mathop{\rm tr}}
\def\Tr{\mathop{\rm Tr}}
\def\Im{\mathop{\rm Im}}
\def\Re{\mathop{\rm Re}}
\def\bR{\mathop{\bf R}}
\def\bC{\mathop{\bf C}}
\def\eq#1{{eq. (\ref{#1})}}
\def\Eq#1{{Eq. (\ref{#1})}}
\def\Eqs#1#2{{Eqs. (\ref{#1}) and (\ref{#2})}}
\def\Eqs#1#2#3{{Eqs. (\ref{#1}), (\ref{#2}) and (\ref{#3})}}
\def\Eqs#1#2#3#4{{Eqs. (\ref{#1}), (\ref{#2}), (\ref{#3}) and (\ref{#4})}}
\def\eqs#1#2{{eqs. (\ref{#1}) and (\ref{#2})}}
\def\eqs#1#2#3{{eqs. (\ref{#1}), (\ref{#2}) and (\ref{#3})}}
\def\eqs#1#2#3#4{{eqs. (\ref{#1}), (\ref{#2}), (\ref{#3}) and (\ref{#4})}}
\def\fig#1{{Fig. (\ref{#1})}}
\def\lie{\hbox{\it \$}} 
\def\partder#1#2{{\partial #1\over\partial #2}}
\def\secder#1#2#3{{\partial^2 #1\over\partial #2 \partial #3}}
\def\bra#1{\left\langle #1\right|}
\def\ket#1{\left| #1\right\rangle}
\def\VEV#1{\left\langle #1\right\rangle}
\let\vev\VEV
\def\gdot#1{\rlap{$#1$}/}
\def\abs#1{\left| #1\right|}
\def\pri#1{#1^\prime}
\def\ltap{\raisebox{-.4ex}{\rlap{$\sim$}} \raisebox{.4ex}{$<$}}
\def\gtap{\raisebox{-.4ex}{\rlap{$\sim$}} \raisebox{.4ex}{$>$}}
\def\lsim{\raise0.3ex\hbox{$\;<$\kern-0.75em\raise-1.1ex\hbox{$\sim\;$}}}
\def\gsim{\raise0.3ex\hbox{$\;>$\kern-0.75em\raise-1.1ex\hbox{$\sim\;$}}}
\def\contract{\makebox[1.2em][c]{
        \mbox{\rule{.6em}{.01truein}\rule{.01truein}{.6em}}}}
\def\half{{1\over 2}}
\def\bel{\begin{letter}}
\def\eel{\end{letter}}
\def\beq{\begin{equation}}
\def\eeq{\end{equation}}
\def\bef{\begin{figure}}
\def\eef{\end{figure}}
\def\bet{\begin{table}}
\def\eet{\end{table}}
\def\bea{\begin{eqnarray}}
\def\ba{\begin{array}}
\def\ea{\end{array}}
\def\bi{\begin{itemize}}
\def\ei{\end{itemize}}
\def\ben{\begin{enumerate}}
\def\een{\end{enumerate}}
\def\ra{\rightarrow}
\def\ot{\otimes}
\def\lrover#1{
        \raisebox{1.3ex}{\rlap{$\leftrightarrow$}} \raisebox{ 0ex}{$#1$}}
%
\def\com#1#2{
        \left[#1, #2\right]}
\def\eea{\end{eqnarray}}
%
\def\bentarrow{\:\raisebox{1.3ex}{\rlap{$\vert$}}\!\rightarrow}
\def\longbent{\:\raisebox{3.5ex}{\rlap{$\vert$}}\raisebox{1.3ex}%
        {\rlap{$\vert$}}\!\rightarrow}
\def\onedk#1#2{
        \begin{equation}
        \begin{array}{l}
         #1 \\
         \bentarrow #2
        \end{array}
        \end{equation}
                }
\def\dk#1#2#3{
        \begin{equation}
        \begin{array}{r c l}
        #1 & \rightarrow & #2 \\
         & & \bentarrow #3
        \end{array}
        \end{equation}
                }
\def\dkp#1#2#3#4{
        \begin{equation}
        \begin{array}{r c l}
        #1 & \rightarrow & #2#3 \\
         & & \phantom{\; #2}\bentarrow #4
        \end{array}
        \end{equation}
                }
\def\bothdk#1#2#3#4#5{
        \begin{equation}
        \begin{array}{r c l}
        #1 & \rightarrow & #2#3 \\
         & & \:\raisebox{1.3ex}{\rlap{$\vert$}}\raisebox{-0.5ex}{$\vert$}%
        \phantom{#2}\!\bentarrow #4 \\
         & & \bentarrow #5
        \end{array}
        \end{equation}
                }
%
%
\def\ap#1#2#3{           {\it Ann. Phys. (NY) }{\bf #1} (19#2) #3}
\def\arnps#1#2#3{        {\it Ann. Rev. Nucl. Part. Sci. }{\bf #1} (19#2) #3}
\def\cnpp#1#2#3{        {\it Comm. Nucl. Part. Phys. }{\bf #1} (19#2) #3}
\def\apj#1#2#3{          {\it Astrophys. J. }{\bf #1} (19#2) #3}
\def\asr#1#2#3{          {\it Astrophys. Space Rev. }{\bf #1} (19#2) #3}
\def\ass#1#2#3{          {\it Astrophys. Space Sci. }{\bf #1} (19#2) #3}

\def\apjl#1#2#3{         {\it Astrophys. J. Lett. }{\bf #1} (19#2) #3}
\def\ass#1#2#3{          {\it Astrophys. Space Sci. }{\bf #1} (19#2) #3}
\def\jel#1#2#3{         {\it Journal Europhys. Lett. }{\bf #1} (19#2) #3}

\def\ib#1#2#3{           {\it ibid. }{\bf #1} (19#2) #3}
\def\nat#1#2#3{          {\it Nature }{\bf #1} (19#2) #3}
\def\nps#1#2#3{          {\it Nucl. Phys. B (Proc. Suppl.) }
                         {\bf #1} (19#2) #3}
\def\np#1#2#3{           {\it Nucl. Phys. }{\bf #1} (19#2) #3}
\def\pl#1#2#3{           {\it Phys. Lett. }{\bf #1} (19#2) #3}
\def\pr#1#2#3{           {\it Phys. Rev. }{\bf #1} (19#2) #3}
\def\prep#1#2#3{         {\it Phys. Rep. }{\bf #1} (19#2) #3}
\def\prl#1#2#3{          {\it Phys. Rev. Lett. }{\bf #1} (19#2) #3}
\def\pw#1#2#3{          {\it Particle World }{\bf #1} (19#2) #3}
\def\ptp#1#2#3{          {\it Prog. Theor. Phys. }{\bf #1} (19#2) #3}
\def\jppnp#1#2#3{         {\it J. Prog. Part. Nucl. Phys. }{\bf #1} (19#2) #3}

\def\rpp#1#2#3{         {\it Rep. on Prog. in Phys. }{\bf #1} (19#2) #3}
\def\ptps#1#2#3{         {\it Prog. Theor. Phys. Suppl. }{\bf #1} (19#2) #3}
\def\rmp#1#2#3{          {\it Rev. Mod. Phys. }{\bf #1} (19#2) #3}
\def\zp#1#2#3{           {\it Zeit. fur Physik }{\bf #1} (19#2) #3}
\def\fp#1#2#3{           {\it Fortschr. Phys. }{\bf #1} (19#2) #3}
\def\Zp#1#2#3{           {\it Z. Physik }{\bf #1} (19#2) #3}
\def\Sci#1#2#3{          {\it Science }{\bf #1} (19#2) #3}
\def\n.c.#1#2#3{         {\it Nuovo Cim. }{\bf #1} (19#2) #3}
\def\r.n.c.#1#2#3{       {\it Riv. del Nuovo Cim. }{\bf #1} (19#2) #3}
\def\sjnp#1#2#3{         {\it Sov. J. Nucl. Phys. }{\bf #1} (19#2) #3}
\def\yf#1#2#3{           {\it Yad. Fiz. }{\bf #1} (19#2) #3}
\def\zetf#1#2#3{         {\it Z. Eksp. Teor. Fiz. }{\bf #1} (19#2) #3}
\def\zetfpr#1#2#3{         {\it Z. Eksp. Teor. Fiz. Pisma. Red. }{\bf #1}
(19#2) #3}
\def\jetp#1#2#3{         {\it JETP }{\bf #1} (19#2) #3}
\def\mpl#1#2#3{          {\it Mod. Phys. Lett. }{\bf #1} (19#2) #3}
\def\ufn#1#2#3{          {\it Usp. Fiz. Naut. }{\bf #1} (19#2) #3}
\def\sp#1#2#3{           {\it Sov. Phys.-Usp.}{\bf #1} (19#2) #3}
\def\ppnp#1#2#3{           {\it Prog. Part. Nucl. Phys. }{\bf #1} (19#2) #3}
\def\cnpp#1#2#3{           {\it Comm. Nucl. Part. Phys. }{\bf #1} (19#2) #3}
\def\ijmp#1#2#3{           {\it Int. J. Mod. Phys. }{\bf #1} (19#2) #3}
\def\ic#1#2#3{           {\it Investigaci\'on y Ciencia }{\bf #1} (19#2) #3}
\def\tp{these proceedings}
\def\pc{private communication}
\def\ip{in preparation}
\relax
\def\e{\mbox{e}}
\def\sgn{{\rm sgn}}
\def\gsim{\;
\raise0.3ex\hbox{$>$\kern-0.75em\raise-1.1ex\hbox{$\sim$}}\;
}
\def\lsim{\;
\raise0.3ex\hbox{$<$\kern-0.75em\raise-1.1ex\hbox{$\sim$}}\;
}
\def\MeV{\rm MeV}
\def\eV{\rm eV}

\thispagestyle{empty}

\begin{titlepage}
\today
\begin{center}
\hfill FTUV/94-15\\
\hfill IFIC/94-14\\
\vskip 0.3cm
\LARGE
{\bf New Supernova Constraints on active-sterile neutrino
 conversions }

\end{center}
\normalsize
\vskip1cm
\begin{center}
{\bf S. Pastor, V. B. Semikoz}
\footnote{On leave from the {\em Institute of the
Terrestrial Magnetism, the Ionosphere and Radio Wave Propagation of the
Russian Academy of Sciences, IZMIRAN,Troitsk, Moscow region, 142092, Russia}.}
{\bf and J. W. F. Valle}
\footnote{E-mail VALLE at vm.ci.uv.es or 16444::VALLE}\\
\end{center}
\begin{center}
\baselineskip=13pt
{\it Instituto de F\'{\i}sica Corpuscular - C.S.I.C.\\
Departament de F\'{\i}sica Te\`orica, Universitat de Val\`encia\\}
\baselineskip=12pt
{\it 46100 Burjassot, Val\`encia, SPAIN}\\
\vglue 0.8cm
\end{center}

\begin{abstract}

We consider active-sterile neutrino conversions in a
supernova in the presence of random magnetic field
domains. For large enough fields the magnetization
of the medium may enhance the active to sterile
\neu conversion rates. Neglecting neutrino transition
magnetic moments we show that for KeV neutrino mass
squared differences these limits may overcome those
that would apply in the isotropic case.

\end{abstract}
\vfill
\end{titlepage}

\newpage

\section{Introduction}

Recently there has been ever growing hints for nonzero
\neu masses from solar and atmospheric neutrino observations
\cite{granadasol} \cite{atm} as well as from the COBE data on
cosmic background temperature anisotropies on large scales
\cite{cobe}. The latter indicate the need for the existence
of a hot dark matter component, contributing about 30\% of the
total mass density \cite{cobe2}.
Simple extensions of the standard electroweak model
that can reconcile all these hints postulate the existence
of a light sterile neutrino $\nu_s$ \cite{DARK92,DARK92B,DARK92D}.
In some of these models such light sterile \neu may play
the role of hot dark matter \cite{DARK92,DARK92D}. Models with
sterile \neus have also been suggested to account for
the possible existence of KeV mass neutrinos \cite{17kev}.
Although the existing data do not support any positive
claim, we stress that such \neus could well exist. Indeed
the only limits on \neu masses of general validity are
those coming from direct laboratory searches. For
example, the cosmological arguments that forbid
\neu masses in the KeV range or above are not
applicable in models with unstable \neus
that decay via majoron emission  \cite{fae}.

So far the most stringent constraints for the neutrino
mass matrix including a fourth neutrino species, $\nu_s$,
are obtained from the nucleosynthesis bound on the maximum
number of extra neutrino species ($\Delta N \leq 0.3$) that
can reach thermal equilibrium before nucleosynthesis and change
the primordially produced helium abundance \cite{Schramm}.
This has been widely discussed in the case of an
isotropic early Universe hot plasma as well as recently
for the case of a large random magnetic field that could
arise from the electroweak phase transition and act as seed
for the galactic magnetic fields \cite{SemikozValle}.

In this letter we would like to consider the effect of
a large random magnetic field in a supernova environment.
The effect of active-sterile neutrino conversions in a
supernova has been discussed both in the isotropic case,
without magnetic field.
In this case stringent constraints can be placed on the
$\nu_a \leftrightarrow \nu_s$ oscillation parameters. For the case
$\Delta m^2 \gsim$ KeV$^2$ one can exclude the range \cite{Maalampi}
\be[v1.1]
2 \times 10^{-2} \gsim \sin^2 2 \theta \gsim 7 \times 10^{-10}
\ee

In a recent paper \cite{Thomson} it was shown that
magnetic fields as strong as $10^{14}$ to $10^{16}$ Gauss
are generated inside a supernova core due to a small scale
dynamo mechanism during the first seconds of neutrino emission.
If the magnetic field is generated after collapse this field
could be viewed as the random superposition of many small
dipoles of strength $10^{14}$ to $10^{16}$ Gauss and size
$L_0 \sim 1$ Km \cite{Thomson}.

In this paper we consider the effect of such huge random
magnetic fields on the active sterile \neu conversions.
We show how such magnetic field can influence the neutrino
spectrum in the supernova medium and thereby modify the neutrino
conversion $\nu_a \ra \nu_s$ rates. We shall
confine ourselves to a given random magnetic field domain
size $L_0$, within which the magnetic field may be taken
as uniform and constant, so that the magnetization of the
medium can be calculated easily.
Although the magnetic field in different domains is
randomly aligned relative to the neutrino propagation
direction, the observable neutrino conversion probabilities
depend on the mean-squared random field via a squared
magnetization value, therefore leading to nonvanishing
averages over the magnetic field distribution.
We apply this to the active-sterile neutrino conversions in
a supernova and show how their effect on the cooling rates
may enable one to place limits that are more stringent than
those that apply in the absence of a magnetic field.

Notice that the effect discussed in our present paper
is more general than that which could also exist due to
nonzero majorana \neus transition magnetic moments
\cite{BFD} as it would exist even if such magnetic
moments are neglected.

\section{Magnetization effects on the active-sterile neutrino
conversions in a supernova}

In this section we consider the active-sterile neutrino
conversions in a supernova core in the presence of a
strong random magnetic field.
We will show here that in a strong magnetic field
supernova constraints on \neu oscillation parameters
can be more stringent than in the isotropic case.
The problem of active to sterile neutrino conversions
in a supernova random magnetic field resembles the one
considered in ref. \cite{SemikozValle}.

Starting from the general equations of motion for a system
of one active (\21 doublet) and one sterile (\21 singlet)
majorana neutrinos \cite{pastor2} one can write the
evolution equation describing their propagation in the
presence of a large random magnetic field, in terms of weak
eigenstates, as
\barray{ll}[v1.20]
i \frac{d}{dt} \Bigl (\matrix{ \nu_a \\ \nu_s }\Bigr ) = \Bigl [
\matrix{ (c^2m^2_1 + s^2m^2_2)/2q + V_{as} + A_{as} & c s \Delta \\
 s c \Delta & (s^2m^2_1 + c^2m^2_2)/2q}\Bigr ] \Bigl (\matrix{ \nu_a \\
\nu_s }\Bigr ),
\ee
where we have denoted by $V_{as}$ the vector part of the
neutrino potential that will describe the active to sterile
conversions, given as
\beq
V_{as} \approx 4\times 10^{-6}\rho_{14}(3Y_e + 4Y_{\ne} - 1)~\MeV,
\label{f6}
\eeq
For a magnetic field in the z-direction the corresponding
axial part $A_{as} (q, B)$ of the neutrino potential that
will describe the active to sterile conversion is given by
\beq
\label{Axial}
A_{as} (q, B) = V^{axial} \frac{q_z}{q}
\eeq
where the term $V_{axial} = \mu_{eff} B$ produced by
the {\sl mean axial current} is proportional to the
magnetization of the plasma in the external magnetic
field
\footnote{We assume this field to be uniform inside a given domain}
and quantity $\mu_{eff}$ is defined, for a degenerate electron gas, as
\beq
\mu_{eff} = \frac{eG_F(-2c_A)p_{F_e}}{\sqrt{2}~2\pi^2}
\approx 3.5 (- 2 c_A) \times
10^{-13}\mu_B\Bigl (\frac{p_{F_e}}{\MeV}\Bigr ).
\label{f2}
\eeq
Here $\mu_B = \mid e\mid /2m_e$ is the Bohr magneton,
$c_A = \mp 0.5$ is the axial coupling specifying the
interaction of charged leptons with \neus in the standard
model (upper sign for \ne and the lower holds for \nm and \nt)
and $p_{F_e}$ is the electron Fermi momentum, given as
\beq
p_{F_e} \approx 320 (Y_e\rho_{14})^{1/3}~\MeV \:.
\eeq
In the above equations q is the \neu momentum,
$m_1$ and $m_2$ are the masses of the \neus, $\theta$
is their mixing angle and we use the standard definitions
$\Delta = \Delta m^2/2q$;
$\Delta m^2 = m^2_2 - m^2_1$;
$c = \cos \theta$, and $s = \sin \theta $.
In addition $\rho_{14}$ is the density of the stellar core in
units $\rho = 10^{14}g/cm^3$, $Y_e$ and $Y_{\nu_e}$ are the
electron and \ne abundances respectively.

Notice that we have neglected the possible effect of
nonzero transition magnetic moments. As we will see,
even in this case, there may be a large effect of the
magnetic field on the conversion rates.

{}From \eq{v1.20} one can easily obtain the probability
$P_{\nu_a \to \nu_s} (t)$ for converting the active
neutrinos $\nu_a$ emitted by the supernova into the sterile
neutrinos, $\nu_s$. In a strong random magnetic field obeying
the condition $\Gamma \gg \Delta_m$ one can write
\be[v1.30]
P_{\nu_a \ra \nu_s} (B,t) \approx
\frac{\Delta^2 \sin^2 2\theta}{2\Delta^2_m}\Bigl (1 -
\exp (-\Delta^2_mt/2\Gamma )\Bigr ),
\ee
which describes the aperiodic behaviour of the active
to sterile \neu conversion. The relaxation time defined as
\be[relax]
t_{relax} = 2\Gamma/\Delta_m^2 = \VEV{\Delta^2_B} L_0/\Delta_m^2
\ee
depends on the mean squared magnetic field parameter
\beq
\langle \Delta^2_B \rangle^{1/2} = {\abs{\mu_{eff}}
\VEV{{\bf B}^2}^{1/2}\over \sqrt{3}},
\label{f1}
\eeq
where $L_0$ is the domain size. In \eq{v1.30} the quantity $\Delta_m$
\beq
\Delta_m = [(V_{as} - \Delta \cos 2\theta )^2 + \Delta^2 \sin^2 2\theta]^{1/2}
\label{f5}
\eeq
is the standard oscillation frequency in the supernova medium
\cite{MSW}.

In order to compare the relaxation time in \eq{relax} with the
mean active \neu collision time $t_{coll} = \Gamma_a(B \neq 0)^{-1}$
we first write the active \neu collision rate in the absence
of magnetic field, obtained in ref. \cite{Maalampi} as
\beq
\Gamma_a (B = 0)\simeq \frac{7.5\times G_F^2\VEV{E}^2}{\pi}\times
\frac{\rho }{m_p}
\label{f3}
\eeq
In the presence of a strong magnetic field $B \gg 2B_c$,
$B_c = m^2_e/e \approx 4.4 \times 10^{13}$ Gauss, the \neu collision
rate may be estimated as \cite{SchrammB},
\be
\Gamma_a(B \neq 0) \lsim 2 B_{14} \Gamma (B = 0)
\ee
As we can see this collision rate could be larger than
$\Gamma_a (B = 0)$ by a factor $2 B_{14}$, where $B_{14}$
denotes the magnetic field strength in units of
$10^{14}$ Gauss.

It is easy to verify that the relaxation time \eq{relax}
can be much larger than the mean active \neu collision time $t_{coll}$.
This allows us, following ref. \cite{SemikozValle}, to average
\eq{v1.30} over collisions so as to obtain
\beq
\VEV{P_{\nu_a \ra \nu_s}  (B)} = \frac{\Delta^2 \sin^2 2 \theta }
{\VEV{\Delta^2_B} 4 \Gamma_a L_0}.
\label{f11}
\eeq
Now we come to the issue of the validity of our
main formula, \eq{v1.30}. For this we evaluate the
damping parameter $\Gamma = \VEV{\Delta^2_B} L_0/2$
using \eq{f1} and \eq{f2}. We find
\beq
\Gamma \approx 4 \times 10^{-6}(Y_e\rho_{14})^{2/3} B_{14}^2~ \MeV .
\label{f8}
\eeq
Thus we see that the requirement that $\Gamma \gg \Delta_m$ is
fulfilled for the case of a strong r.m.s. magnetic field obeying
\footnote{In this estimate we have assumed $\Delta_m \sim V_{as}$}
\beq
B_{14} \gg 1.6 \times \rho_{14}^{1/6}\times \frac{(3Y_e + 4Y_{\nu e}
- 1)^{1/2}}{Y_e^{1/3}}
\label{f9}
\eeq
It has been suggested that very large random magnetic fields
as strong as $10^{14}-10^{16}$ Gauss may in fact be generated
inside a supernova core or a nascent neutron star due to a small
scale dynamo mechanism that could take place after collapse during
the first few seconds of neutrino emission \cite{Thomson}.
These authors indeed suggested that such field could behave
like a random superposition of many small dipoles of strength
$\gsim 10^{14}$ Gauss and size $L_0 \sim 1$ Km.

Note finally that in the opposite case of weak magnetic fields
$\Gamma \ra 0$ the probability \eq{v1.30} coincides with the
averaged non-resonant MSW result \cite{MSW}
\beq
\label{relation}
\VEV{P_{\nu_a \ra \nu_s} (B \ra 0)} =
\frac{\Delta^2 \sin^2 2\theta}{2 \Delta^2_m}
\equiv \frac{\sin^2 2\theta_m}{2}
\eeq
used in ref. \cite{Maalampi} in order to derive constraints
on \neu oscillation parameters in the isotropic case.

It will be convenient, in analogy to the isotropic result
$P_{\nu_a \ra \nu_s} (B \ra 0) = \sin^22\theta_m/2$,
to rewrite \eq{f11} as
$$
\VEV{P_{\nu_a \ra \nu_s}  (B \neq 0)} = \frac{\sin^2 2\theta_B}{2},
$$
where we define the mixing angle in the presence of the
magnetic field via
\beq
\sin^2 2 \theta_B = \frac{\Delta^2 \sin^2 2\theta}
{2 \VEV{\Delta^2_B} \Gamma_a L_0}
=  \frac{x}{2} \sin^2 2\theta_m~,
\label{sinmag}
\eeq
Notice that for a very strong magnetic field, for instance,
a r.m.s. field $B_{14} \sim 10^2$ the parameter
$x = \Delta^2_m/ 2 \Gamma \Gamma_a (B \neq 0)$
becomes less than unity
\beq
x \approx \frac{1.5 \times 10^6 \rho_{14}^{1/3}(3Y_e + 4Y_{\nu e} - 1)^2}
{B_{14}^3Y_e^{2/3}\VEV{E^2_{100}}} \leq 1,
\label{argument}
\eeq
if the abundance factor (for $\VEV{E_{100}^2} = 1$, i.e. for
100 MeV \neu energies) is less than
\beq
\abs{3Y_e + 4Y_{\ne} - 1} \leq  Y_e^{1/3}\rho_{14}^{-1/6}
\label{aban1}
\eeq
If, on the other hand, the argument in \eq{argument} is much
less than unity, $x \ll 1$, the probability \eq{v1.30},
averaged over fast collisions, reproduces the result in
\eq{f11}. Notice that the mixing angle in the presence of
the magnetic field is smaller than with B=0, as the energy
difference between the two diagonal entries in \eq{v1.20}
increases due to the presence of the extra axial term.
Let us now turn to astrophysics.

\section{Supernova Constraints}

There are two ways to place constraints on \neu oscillation
parameters using astrophysical criteria, depending on the relative
value of the sterile neutrino effective mean free path
$l_s = \Gamma_s^{-1} = [P (\nu_a \ra \nu_s) \Gamma_a]^{-1}$
and the core radius $R_{core}$, where the effective sterile
\neu production rate $\Gamma_s$ is the product of the active
\neu production rate (which in our estimate we take to be the
same as $\Gamma_a$) with the active to sterile conversion probability.

If the trapping condition $l_s \leq R_{core}$ is fulfilled,
the \ns are in thermodynamical equilibrium with the medium and,
due to Stefan-Boltzman law, the ratio of the sterile neutrino
luminosity to that of the ordinary neutrinos
$$
\frac{Q_s}{Q_a}\simeq \Bigl (\frac{T(R_s)}{T(R_a)}\Bigr )^4\Bigl (
\frac{R_s}{R_a}\Bigr )^2\simeq \Bigl (\frac{\Gamma_a}{\Gamma_s}\Bigr )^
{1/2}
$$
does not depend on $\Gamma_a$.
In this first regime one considers surface thermal neutrino
emission and sets the conservative limit $(Q_s/Q_a)_{max} \gsim 10$
in order to obtain the excluded region of \neu parameters.
In the isotropic case one has
\beq
\frac{Q_s}{Q_a} \simeq (\frac{\sin^2 2\theta_m}{2})^{-1/2}~.
\label{limit1}
\eeq
which gives the excluded region \cite{Maalampi}
\beq
\sin^2 2 \theta_m \lsim 2 \times 10^{-2}~.
\label{limit2}
\eeq

Another complementary constraint can be obtained from the
requirement that in the non-trapping regime the sterile
neutrino can be emitted from anywhere inside the star
volume with a rate
\beq
\frac{dQ(B=0)}{dt} \simeq \frac{4}{3}\pi R^3_{core}n_{\nu e}\Gamma_s\VEV{E_s}
\simeq 1.4\times 10^{55}\sin^2 2\theta_m \rm{\frac{J}{s}}
\label{limit3}
\eeq
which should not exceed the maximum observed integrated neutrino
luminosity. For instance, for the case of SN1987A, this is
$\sim 10^{46}~$ J. For the isotropic case the corresponding
collision rate should be less than $10^{46}~J/s$, so that
one obtains the excluded region \cite{Maalampi}
\beq
\sin^2 2\theta_m \gsim 7 \times 10^{-10}
\label{limit4}
\eeq
We now move to the case of strong magnetic field and
how the constraints that can be obtained for this case
compare with those of the isotropic case. For $B \neq 0$
we use the known estimate for the active neutrino collision
rate $\Gamma_a(B\neq 0)\leq 2B_{14}\Gamma_a(B=0)$ \cite{SchrammB}
and the relationship between the corresponding conversion
probabilities in order to obtain $\frac{dQ(B \neq 0)}{dt}$.
The ratio of sterile neutrino volume energy losses in the
presence and absence of magnetic field may be easily obtained as
\beq
\frac{dQ(B = 0)/dt}{dQ(B \neq 0)/dt} \sim \frac{1}{x B_{14}} ,
\label{limit6}
\eeq
where $x$ is the small parameter in \eq{argument}. From the
last inequality we can find a region of abundances \eq{aban1}
where our result for the conversion probability \eq{f11} is
valid ($x\ll 1$) so that we obtain the excluded region
\beq
\sin^2 2\theta_m \gsim \frac{7 \times 10^{-10}}{x B_{14}} \:.
\label{small}
\eeq
Note that this constraint on the \neu parameters
can be more stringent than that of \eq{limit4}.

In an analogous way we obtain the ratio of luminosities
$Q_s/Q_a$ in the trapping regime (this is independent of
the active neutrino collision rate $\Gamma_a(B)$) and find
that $\sin^2 2\theta_B \leq 2 \times 10^{-2}$ leading, from
\eq{sinmag}, to a new excluded region of the active-sterile
\neu mixing angle
\beq
\label{}
\sin^2 2\theta_m \lsim \frac{4 \times 10^{-2}}{x}
\eeq
Note that this constraint on the \neu parameters
can be more stringent than that of \eq{limit2}.
In particular, for a supernova with strong magnetic field
it is possible to exclude all region of large mixing angles,
if the parameter $x$ in \eq{argument} is $x \leq 0.04$.
This will be realized for a r.m.s. field $B_{14} \sim 10^2$
\cite{Thomson} and mean sterile neutrino energy $\VEV{E_{100}^2}=1$
if the abundance parameter is less than
\beq
\mid 3Y_e + 4Y_{\nu e} -1\mid \leq .2 \times
Y_e^{1/3}\rho_{14}^{-1/6}~.
\label{abun}
\eeq
This condition can indeed be realized for a stage of
supernova after bounce \cite{Maalampi,Notzold}. Moreover,
this assumption is not crucial for us, in contrast to the
case of resonant neutrino spin-flip due to a neutrino
magnetic moment considered in ref. \cite{Voloshin}.

We now summarize this discussion with one example.
Our goal is to display the above constraints in terms of
the parameters $\Delta m^2 \sin^2 2\theta$. In order to do
this we use \eq{relation} to relate the \neu
parameters in matter with those in vacuum.
In Figure 1 we show the constraints on \neu parameters
that follow from supernova cooling rates in the presence
of a strong random magnetic field. The region of \ne to
\ns parameters
\footnote{Similarly one may derive constraints on
\nm or \nt to \ns conversions}
above the dotted line is excluded by supernova cooling
rates for aperiodic conversion in a random magnetic field
$B = 10^{16}$ Gauss. In contrast, in the isotropic case,
for the same parameters only the points between the solid
lines would be excluded.

\section{Discussion and conclusions}

The possible existence of huge random magnetic fields
generated during the first few seconds of neutrino emission
modifies the \neu spectrum in a supernova due to the
magnetization of the medium.

Averaging the differential equation describing
the evolution of the $ \nu_a \to \nu_s$ conversion
probabilities over the random magnetic field distribution
we find a nonvanishing mean squared field effect which
drastically changes the conversion rates with respect
to those of the isotropic case. This may be used to
derive new and more stringent constraints on the
active-sterile neutrino oscillation parameters than
in the isotropic case without random magnetic field.

In the presence of a large magnetic field
the active to sterile \neu conversion probability
is suppressed relative to that in the isotropic case due to
the larger energy difference between the two diagonal
entries in the \neu evolution hamiltonian \eq{v1.20}
caused by the presence of the extra axial term.
Note, however the sterile \neu production rate could
be larger in this case due to effect of the large magnetic
field.

On the other hand the ratio of active and sterile \neu
thermal luminosities does not depend on the active \neu
production rate. However, the smaller the conversion
probability the larger the sterile \neu effective mean
free path and therefore they can leave the star more easily
than in the isotropic case. This leads to the possible
exclusion of the complete large mixing angle region,
al illustrated in Fig. 1.

Notice that, although majorana \neus could have nonzero
transition magnetic moments \cite{BFD}, we have neglected
them in our present discussion.
As we have seen, even in this case, there may be a large
effect of the magnetic field on the \neu conversion rates.
The effect which we have found is therefore of more
general validity than that which could be ascribed
to nonzero electromagnetic moments. The latter are
expected to be small in the simplest extensions of
the standard model.

\vfill
\noi

{\bf Acknowledgements}
This work was supported by DGICYT under grant number
PB92-0084 and partially by the EEC under grant number
CHRX-CT93-0132, and by a senior researcher NATO fellowship
(V.S.). We thank Armando Perez for pointing out to us the
paper in ref. \cite{Thomson} and also Juan Antonio Miralles
for fruitful discussions.

\vskip 1truecm

\newpage
\begin{center}
\vskip 5cm
{\bf Figure Caption}
\end{center}
\vskip2cm

{\bf Fig.1.} \\

Constraints on active to sterile \neu conversions
from a supernova with a strong random magnetic field.
The region of \ne to \ns $\Delta m^2 \sin^2 2\theta$
parameters above the dotted line is excluded by supernova
cooling rates for aperiodic conversion in a strong random
magnetic field $B = 10^{16}$ Gauss. Here we assume
$Y_e \simeq 0.3$, $Y_{\nu e} \simeq 0.045$ and core
density $\rho = 8 \times 10^{14}$ g/cm$^3$.
For comparison the region between the solid lines
would be excluded for the same parameters for the
case of $\nu_a \leftrightarrow \nu_s$ oscillations
in the isotropic case B=0.


\newpage

\begin{thebibliography}{99}

\bibitem{granadasol}
T. Kirsten, \pl{B314}{93}{445};
Kamiokande, SAGE, GALLEX collaboration talks
at {\sl XV Int. Conference on Neutrino Physics
and Astrophysics}, \nps{31}{93}{105-124};
 J. R.~Davis in {\sl Proceedings of the 21th
International Cosmic Ray Conference,  Vol. 12},
ed.\  R.~J. Protheroe (University of Adelaide Press, 1990) p. 293.

\bibitem{atm}
Kamiokande collaboration, \pl{B205}{88}{416},
\pl{B280}{92}{146} and \pl{B283}{92}{446} ;
IMB collaboration, \pr{D46}{92}{3720};
see also the proceedings of {\it Int. Workshop on
\nm/\ne problem in atmospheric \neus} ed. V. Berezinsky
and G Fiorentini, Gran Sasso, 1993.

\bibitem{cobe}
G.~F. Smoot et~al., \apj{396}{92}{L1-L5}.

\bibitem{cobe2}
E.L.~Wright et al., \apj{396}{92}{L13};
M.~Davis, F.J.~Summers, and D.~Schagel, \nat{359}{92}{393};
A.N.~Taylor and M.~Rowan-Robinson, \ib{359}{92}{396};
R.K.~Schaefer and Q.~Shafi, \nat{359}{92}{199};
J.A.~Holtzman and J.R.~Primack, \apj{405}{93}{428};
A.~Klypin et al., \apj{416}{93}{1}

\bibitem{DARK92}
J.~T. Peltoniemi, D.~Tommasini, and J. W. F. Valle,
\pl {B298}{93}{383}

\bibitem{DARK92B}
J.~T. Peltoniemi, and J. W. F. Valle, \np{B406}{93}{409};
D.O.~Caldwell and R.N.~Mohapatra, \pr{D48}{93}{3259}

\bibitem{DARK92D}
E. Akhmedov, Z. Berezhiani, G. Senjanovic and Z. Tao, \pr{D47}{93}{3245};
J. T. Peltoniemi, \mpl{A38}{93}{3593}

\bibitem{17kev}
L. Bento and J.W.F.Valle, \pl{B264}{91}{373};
A. Smirnov and J.W.F.Valle, \np{B375}{92}{649};
J. Peltoniemi, A. Smirnov and J.W.F.Valle, \pl{B286}{92}{321}

\bibitem{fae}
For a recent review see J W F Valle, in {\it Gauge Theories and the
Physics of Neutrino Mass}, \ppnp{26}{91}{91-171} and references therein.

\bibitem{Schramm}
T.Walker, G.Steigman, D.N.Schramm, K.Olive, and H.Kang,
\apj{376}{91}{51}.

\bibitem{SemikozValle}
V. Semikoz and J.W.F.Valle, Valencia Preprint FTUV/94-05

\bibitem{Maalampi}
K. Kainulainen, J. Maalampi and J.T. Peltoniemi, \np{B358}{91}{435};
G. Raffelt and G. Sigl, Astroparticle Physics 1 (1993) 165.

\bibitem{Thomson}
C. Thomson and R.C. Dunkan, \apj{408}{93}{194}.

\bibitem{BFD}
J. Schechter and  J. W. F. Valle, \pr{D24}{81}{1883};
\pr{D25}{82}{283}

\bibitem{pastor2}
S. Pastor, V. Semikoz and J.W.F.Valle, \ip

\bibitem{MSW}
M. Mikheyev, A. Smirnov, \sjnp{42}{86}{913};
L. Wolfenstein, \pr {D17}{78}{2369};\ib{D20}{79}{2634}.

\bibitem{SchrammB}
B. Cheng, D.N. Schramm and J.W. Truran, \pl{B316}{93}{521}.

\bibitem{Notzold}
D. Notzold \pr{D38}{88}{1658}

\bibitem{Voloshin}
M.B. Voloshin, \pl{B209}{88}{360}.

\end{thebibliography}
\end{document}